\documentclass[structabstract]{aa}  
\usepackage{graphicx}
\usepackage{txfonts}
\newcommand{\alfa}{$\alpha$}
\newcommand{\alfaFe}{[$\alpha$/Fe]}
\newcommand{\meta}{[M/H]}
\newcommand{\T}{$T_{\rm eff}$}
\newcommand{\g}{log($g$)}
\begin{document}
   \title{The AMBRE project: A new synthetic grid of high-resolution FGKM stellar spectra}

   \author{P. de Laverny\inst{1}
          \and
          A. Recio-Blanco\inst{1}
          \and
          C.C. Worley\inst{1}
          \and
          B. Plez\inst{2}
          }

   \institute{Laboratoire Lagrange (UMR7293), Universit\'e de Nice Sophia Antipolis, CNRS, Observatoire de la C\^ote d'Azur, BP 4229,
 F-06304 Nice cedex 4, France\\
              \email{laverny@oca.eu}
         \and
Laboratoire Univers et Particules de Montpellier, Universit\'e Montpellier II, CNRS, 34095 Montpellier Cedex 05, France
             }

   \date{Received 2 april 2012; accepted 6 may 2012}

   \abstract
    {Large grids of synthetic spectra covering a widespread range of stellar parameters are mandatory for different
stellar and (extra-)Galactic physics applications. For instance, such large grids can be used for 
the automatic parametrisation of stellar spectra such as that performed within the AMBRE project for which the main goal is the stellar atmospheric
parameters determination for the few hundreds of thousands of archived spectra of four ESO spectrographs.}   
    {To fulfil the needs of AMBRE and future similar projects, we have computed a grid of synthetic spectra over the whole optical domain
for cool to very cool stars of any luminosity (from dwarfs to supergiants) with metallicities varying from 10$^{-5}$ to 10 times the solar metallicity, and considering large variations in the chemical content of the \alfa -elements.}
    {For these spectrum computations, new generation MARCS model atmospheres and the Turbospectrum 
code for radiative transfer have been used. We have also taken into account as complete as possible atomic and molecular linelists 
and adopted, in the spectral synthesis, the same physical assumptions and input data as in the MARCS models. This allowed us to
present a grid with a high consistency between the atmosphere models and the synthetic spectra.}
    {A new grid of 16\,783 high resolution spectra over the wavelength range 3\,000 to 12\,000~\AA \, has been computed with a spectral resolution always larger than 150\,000.
Normalised and absolute flux versions are available over a wide range of stellar atmospheric parameters for
stars of FGKM spectral types. The covered parameters are 2\,500~K$\leq$\T$\leq$8\,000~K, -0.5$\leq$\g$\leq$5.5~dex and
-5.0$\leq$\meta$\leq$+1.0~dex. Moreover, for each combination of these stellar parameters, five different values
of the enrichment in \alfa-elements have been considered (0.0, $\pm$0.2~dex and $\pm$0.4~dex around the standard values).
This library is thus relevant for any stellar type and luminosity class, present in old and intermediate-age stellar
populations from extremely metal-poor to metal-rich in content.
This grid is made publicly available through the POLLUX database (about 50\% of the spectra are already included
in this database) 
and in FITS format upon request to the authors.}
  {} 


   \maketitle
%

\section{Introduction}
AMBRE is a project defined between the Observatoire de la
C\^ote d'Azur and the European Southern Observatory (ESO). 
Its main goal is the automated stellar parametrisation
of the FEROS, HARPS, UVES and GIRAFFE archived spectra.
This parametrisation of more than 300\,000 spectra  will be made
publicly available through the ESO archive. Furthermore, AMBRE offers
the opportunity to test on a substantial amount of real spectra the
performances of parametrisation algorithms such as MATISSE (\cite{Recio})
or DEGAS (\cite{Kordopatis}) which have been specifically developed for the analysis
of the Radial Velocity Spectrometer (RVS) data collected by the ESA/Gaia mission.
We refer the reader to \cite{Worley} for a detailed description of AMBRE and the complete analysis 
of the FEROS spectra.

Within the AMBRE project, the adopted algorithms for
the automatic parametrisation of stellar spectra rely on a library of reference 
spectra to derive the atmospheric parameters and the chemical abundances.
Preparing such a spectra grid, covering a wide range of parameters
is mandatory in order to allow the rapid analysis with adapted algorithms of large quantities of observed spectra
from any stellar type, such as those provided by current (RAVE, SEGUE, Gaia/ESO Survey.. )
and future (Gaia/RVS) spectroscopic surveys. Therefore,
when planning AMBRE, libraries of observed (empirical) and/or computed (synthetic) stellar spectra 
were investigated. We ultimately concluded that there is the lack of an extensive library of stellar spectra
with the following properties: (i)
coverage of a wide range of atmospheric parameters,
(ii) coverage of a very large spectral domain in the optical, (iii) high spectral resolution, (iv)
well-established atmospheric parameters, (v) inclusion of chemical abundance variations.

Indeed, although important progress has been made over the last few years to build
high-quality empirical libraries of stellar spectra (see, in particular, the
last release of the CFLIB library by \cite{Wu}), observed spectra libraries still often suffer from
severe shortcomings. For instance,
most of them rely on rather low resolution spectra (CFLIB or MILES, \cite{MILES}) 
or rather restricted wavelength coverage (as the ELODIE library, \cite{Elodie})
and/or suffer from flux calibration uncertainties. Furthermore, the atmospheric parameters attached to these compiled
stellar spectra may be subject to caution due to the lack of determination of some parameters, 
different adopted bibliographic sources and/or method of determination (see \cite{PASTEL}). However we
point out the important effort conducted recently by \cite{Wu} to obtain a homogeneous set of
stellar atmospheric parameters. But this new CFLIB library still suffers from a rather incomplete
coverage of the parameter space (particularly for cool and/or metal-poor stars). Furthermore, no
information on the chemical abundance of \alfa \, elements is provided (to our knowledge, all
available empirical libraries provide only the main three, in the best cases, atmospheric parameters, i.e. the effective temperature,
the surface gravity and the mean metallicity). All of these reasons prevented the use of these empirical libraries
within AMBRE because, as mentioned previously, the main goal of AMBRE is to derive, from high-resolution spectra covering very different regions
of the whole optical domain, the atmospheric parameters together with the \alfaFe \, enrichment of the target stars
for the purposes of galactic archaeology.

Regarding the synthetic spectra libraries that are available, the situation was also not completely satisfactory.
\cite{Munari} present an extensive grid of cool to very hot stars based on Kurucz/ATLAS9 model
atmospheres and the SYNTHE spectral synthesis code (see also their Tab.~1 for a comparison of 
the principal characteristics of most publicly available libraries before 2005).
Although the number of available spectra is impressive (because the authors considered
different values of microturbulent and rotational velocities, spectral resolution and sampling law),
these spectra actually correspond to just a few thousand combinations of \T , \g , \meta \, values. Furthermore,
the largest spectral resolution ($R\sim$20\,000) is too low for the analysis of the ESO
spectra and, as for the empirical libraries, no variations of the \alfaFe \, enrichment
have been considered (two values of \alfaFe \, were computed at three distinct metallicities, only). Finally,
the metallicity range is also too small for general parametrisation (-2.5 $\leq $ \meta $\leq$ 0.5).
Another synthetic high-resolution spectra grid has been presented by \cite{Coelho}, again
based on Kurucz/ATLAS9 models. This grid covers the whole 
optical domain but, once again, the adopted variations in \alfaFe \, are still not satisfactory
(only \alfaFe = 0 and 0.4~dex are considered for every models) for the automatic parametrisation purposes
for which this chemical index is searched.
Finally, \cite{Kirby} published a grid very similar to our needs in terms
of atmospheric parameter ranges but, unfortunately, it is dedicated to the red part
of the optical domain ($\lambda>6\,300$~\AA). These spectra are computed from
a new grid of Kurucz/ATLAS9 atmosphere models and the MOOG code. There is however a
lack of molecular lines considered in the spectral synthesis. Only the CN, C$_2$ and MgH molecules were
considered (partly because
the number of TiO lines to be considered led to prohibitive computational times).
Therefore, the provided spectra for cool and/or metal-rich stars might not be well modelled
in this grid. 

For all these reasons, we decided to compute a new synthetic grid of stellar spectra
in order to fulfil the AMBRE needs.
The goal of this paper is to present and make such a grid publicly available. 
The computed spectra cover the whole optical spectral range
(from 3\,000 to 12\,000~\AA ) at high resolution ($R > 150\,000$) in order to be used for the analysis
of most spectrograph data. They correspond to FGKM stars of any luminosity (except White Dwarfs)
with a very wide range of metallicities and \alfa-element abundance variations at any metallicity.
Furthermore, it is the first grid based exclusively on MARCS model atmospheres and associated
tools to compute the synthetic spectra
(with therefore the same physical assumptions and ingredients, such as the extensive molecular linelists). So far, it is also
the most extensive grid by about one order of magnitude in the number of available
spectra of cool stars 
(mostly because we span cooler temperatures, lower metallicities and several variations in \alfaFe ).
This grid can easily be used as reference for the automatic determination of atmospheric parameters
(as carried out in the AMBRE project) or for building composite synthetic stellar populations for extragalactic
studies.
In particular, the simulation of the integrated spectra of dwarf galaxies as those found in the Milky Way vicinity
with stars having large or small \alfaFe \, values (Frebel
et al. 2010; Letarte et al. 2010) requires such a library with a
broad range of \alfaFe \, chemical enrichments. This is also mandatory for future and present galactic 
spectroscopic surveys looking for accreted halo stars.

This paper is organised as follow:
Sect.~ 2 is devoted to the description of the methods, assumptions and input data adopted for 
generating the stellar spectra. The AMBRE grid of synthetic spectra is then presented 
in Sect.~3. Sect.~4 \& 5 discuss some properties and limitations of this grid. We present in Sect.~6 how these spectra 
can be retrieved. The conclusions are given in Sect.~\ref{conclu}.

\section{Computation of the synthetic spectra}
The AMBRE project is primarily devoted to the analysis of FGKM stars, i.e. the cool stars
that are the main targets of the MARCS model atmospheres for almost 40 years (\cite{G75}).
Therefore, the presented grid is based on the last generation of MARCS model atmospheres 
presented in \cite{G08}. The synthetic spectra have been generated in a consistent
way (as far as possible) with these models.
The spectra were indeed computed with the Turbospectrum code for spectral synthesis V9.2 (\cite{Alvarez} with subsequent improvements through the years by B. Plez). This code uses
the same routines and input data as the MARCS code. We remind that MARCS models are computed 
assuming one-dimensional plane-parallel
or spherical geometry (depending on the gravity, see below),
hydrostatic and local thermodynamic equilibria
and convection is treated following the mixing-length theory.

We have seen that, so far, all the other available grids of synthetic spectra
are based on Kurucz/ATLAS9 model atmospheres. \cite{G08} present a comparison
of their new MARCS models with the ATLAS9 ones (see, also, Plez 2011). They report a rather
good agreement in the temperature structures for solar-type or cooler dwarf stars of different metallicities. 
However, larger discrepancies appear at lower \T \, and/or lower \meta . Furthermore, for giants, 
MARCS models appear slightly cooler in the upper layers (probably due to the
improved molecular data collected by the MARCS group over the years). 
It is therefore expected that the presented AMBRE spectra grid may represent 
an important improvement towards the modelling of cool stars of any luminosity (from dwarfs to supergiants)
and metallicity.

\subsection{Atomic and molecular line opacities}

\begin{table*}[ht]
\caption{Isotopic compositions adopted in the calculations of the molecular equilibrium (see, for references,
the Turbospectrum code).}
\label{Tab1}
\begin{tabular}[]{clrclrclrclrclrc}
   \hline
   \hline
   \noalign{\smallskip}
    Element & \multicolumn{14}{c}{Isotopes}  \\
   \noalign{\smallskip}
   \hline
   \noalign{\smallskip}
   C  & $^{12}$C  & 98.90\%  & & $^{13}$C &   1.10\%  &  &          &           &  &          &           &  & & \\
   N  & $^{14}$N  & 99.634\% & & $^{15}$N &   0.366\%&  &          &           &  &          &           &  & & \\
   O  & $^{16}$O  & 99.762\% & & $^{17}$O &   0.038\%&  & $^{18}$O &   0.20\% &  &          &           &  & & \\
   Mg & $^{24}$Mg & 78.99\%  & & $^{25}$Mg&   10.00\%&  & $^{26}$Mg&   11.01\%  &  &          &           &  & & \\
   Si & $^{28}$Si & 92.23\%  & & $^{29}$Si&   4.67\% &  & $^{30}$Si&  3.10\%  &  &          &           &  & & \\
   Ca & $^{40}$Ca & 96.941\% & & $^{42}$Ca&   0.647\%&  & $^{43}$Ca&  0.135\%&  & $^{44}$Ca&  2.086\%&  & $^{48}$Ca& 0.187\%  \\
   Ti & $^{46}$Ti & 8.00\%   & & $^{47}$Ti&   7.30\%  &  & $^{48}$Ti&  78.30\% &  & $^{49}$Ti&  5.50\%  &  & $^{50}$Ti& 5.40\%\\
   V  & $^{50}$V  & 0.25\%   & & $^{51}$V &   99.75\%&  &          &           &  &          &           &  & & \\ 
   Fe & $^{54}$Fe & 5.80\%   & & $^{56}$Fe&   91.72\%&  & $^{57}$Fe&  2.20\%  &  & $^{58}$Fe&  0.28\% &  & & \\
   Zr & $^{90}$Zr & 51.46\%  & & $^{91}$Zr&   11.22\%&  & $^{92}$Zr&  17.15\%&  & $^{94}$Zr&  17.38\%&  & $^{96}$Zr&  2.80\%\\
  \noalign{\smallskip}
   \hline
\end{tabular}
\end{table*}

In the spectrum calculations a number of specific line lists for hydrogen, metals, and
molecules covering the 3000-12\,000~\AA \, spectral range were considered. 

Atomic line data were recovered from 
the Vienna Atomic Line Database (content of august 2009; \cite{Kupka}).
Only lines from neutral or singly ionised atoms were kept (which is a reasonable
assumption for the considered effective temperatures). This led
to a total of about 215\,000 atomic lines from 76 neutral metals and
60 singly ionised metals (among those lines, about 35\,100 of them are transitions
of Fe~I and 33\,600 of Fe~II).
Due to the huge amount of adopted lines over the wide spectral range,
no correction or astrophysical calibration of the atomic data were performed
(the completeness and the quality of the VALD linelists could
be questionned although this database is continuously
improved thanks to huge efforts conducted by several groups). 
Therefore, we warn future users of this synthetic spectra grid
that it is not optimised for fine analysis or abundance determination from fitting 
individual spectral lines.
As in \cite{G08}, hydrogen self broadening is estimated
following Barklem et al. (2000a) and damping constants for many strong metallic
lines are calculated using recent data (Barklem et al. (2000b) and related
references)\footnote{One can expect some differences (particularly in the H-line profiles)
in our computed fluxes when compared to
previous grids since such improvements have never been considered up to now by other groups
computing large spectra grids.}. Other lines are broadened with the classical Uns\"old theory
with an assumed enhancement factor for the Van der Waals broadening set to 2.5 except for NaI, SiI, CaI
and FeI for which the enhancement factor was set to 2.0, 1.3, 1.8 and 1.4, respectively
(see \cite{G08}).

The considered molecular lines come from the CH, NH, OH, 
MgH, SiH, CaH, FeH, C2, CN, TiO,
 VO, and ZrO molecules with their corresponding
isotopic variations. 
In these calculations, we disregarded CO lines since they
appear beyond the IR wavelength limit of the present grid at $\lambda >$ 1.5~microns.
We however point out that this molecule together with several other molecular species have been 
considered in the calculation of the chemical equilibrium (see the complete list in \cite{G08}).
In total, 41 different species (see Tab~\ref{Tab1} for the adopted isotopic ratios)
consisting of more than 20 millions molecular lines were taken into account (half of these
lines being TiO transitions and one fourth belong to the ZrO molecule). 
These molecular line lists have been compiled by B. Plez
and are identical to those used by \cite{G08}
(see their Tab.~2 for a list of
references). 
We point out that, for molecular lines, the damping was not taken into account for the same
reasons as in the MARCS models.

\subsection{Selection of MARCS model atmospheres}

\begin{table*}[ht]
\caption{Atmospheric parameters of the AMBRE grid. The description for the other parameters adopted in the spectra
calculation is detailed in 
Sect.~2 \& 3. We also recall that (i) for each set of (\T, \g\, \meta) values, 5 spectra with different \alfaFe \, enrichments have been computed, and that (ii) some models are missing (approach to the Eddington
flux limit and/or poor convergence).}
\label{TabGrid}
\begin{tabular}[]{llllll}
   \hline
   \hline
   \noalign{\smallskip}
    Atmospheric Parameter (units) & Minimum & Maximum & Step & & Number of possible values\\
   \noalign{\smallskip}
   \hline
   \noalign{\smallskip}
\T \, (in K)        & 2500 & 8000 & 200  & for \T $\leq$ 3900     & 25 \\
                    &      &      & 250  & for \T $\geq$ 4000     &    \\
\noalign{\smallskip}
\g \, (in cm/s$^2$) & -0.5 & +5.5  & 0.5  &                        & 13 \\
\noalign{\smallskip}
\meta \, (in dex) & -5.0 & +1.0   & 0.25 & for \meta $\geq$ -1.0  & 15 \\
                  &      &        & 0.5  & for -3.0 $\leq$ \meta  $<$ -1.0 &    \\
                  &      &        & 1.0  & for \meta  $<$ -3.0 &    \\
\noalign{\smallskip}
\alfaFe  \, (in dex) & -0.4 & +0.8 & 0.2  &                        & 5 for a given \meta  \\
\noalign{\smallskip}
   \hline
\end{tabular}
\end{table*}
We recovered 3\,358 MARCS model atmospheres (version of September 2009; only
converged models were kept). 
The two main atmospheric parameters of the selected models (see Tab.~\ref{TabGrid}) are the 
effective temperature (\T) and the stellar surface gravity (\g).
The effective temperatures vary between 2\,500~K and 8\,000~K (step of 200~K below 4\,000~K and 250~K above)
and the logarithm of stellar surface gravities from -0.5 to 5.5~dex (step of 0.5~dex), $g$ in units of cm/s$^2$.

\begin{figure*}[ht]
\includegraphics[width=9.2cm,height=9.2cm]{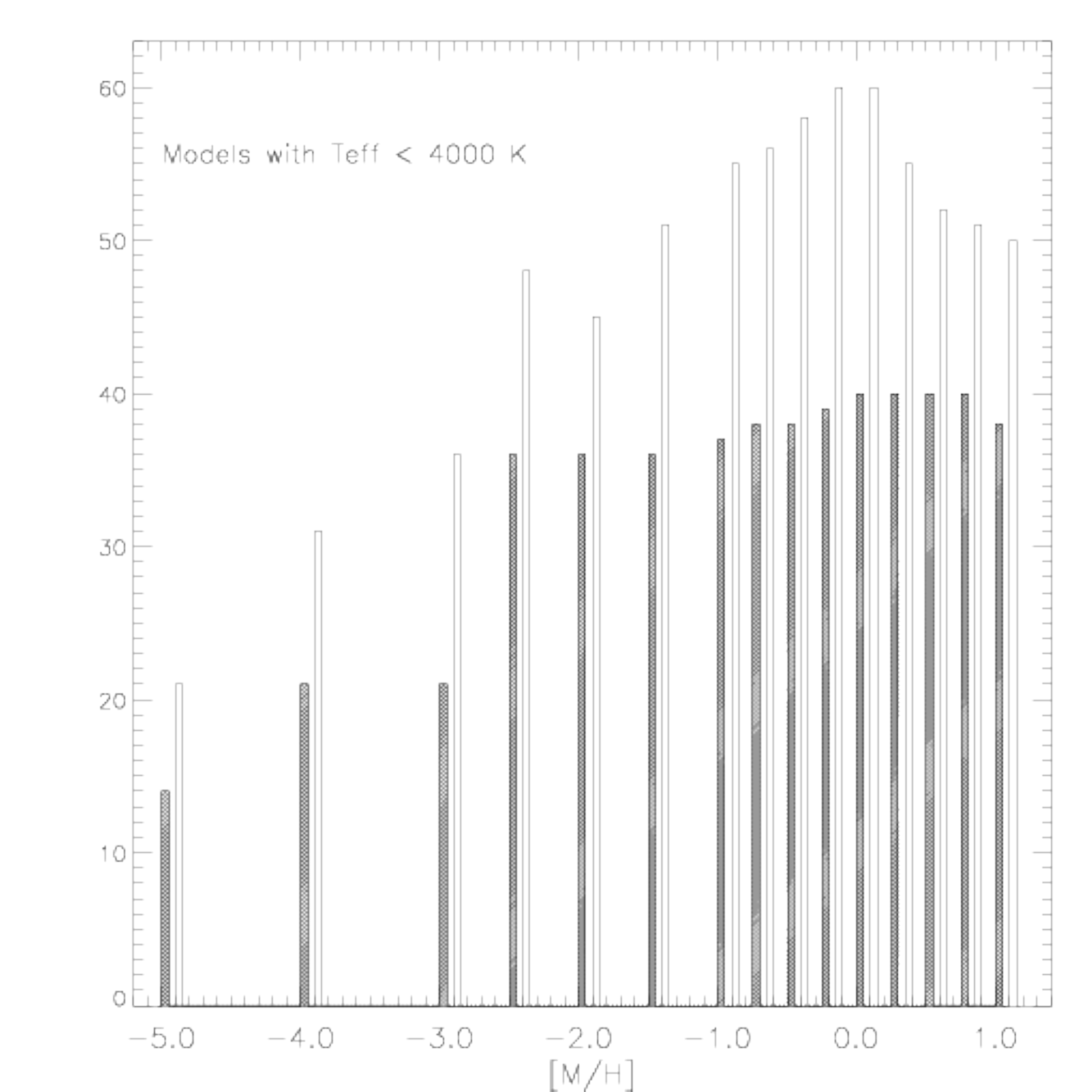}
\includegraphics[width=9.2cm,height=9.2cm]{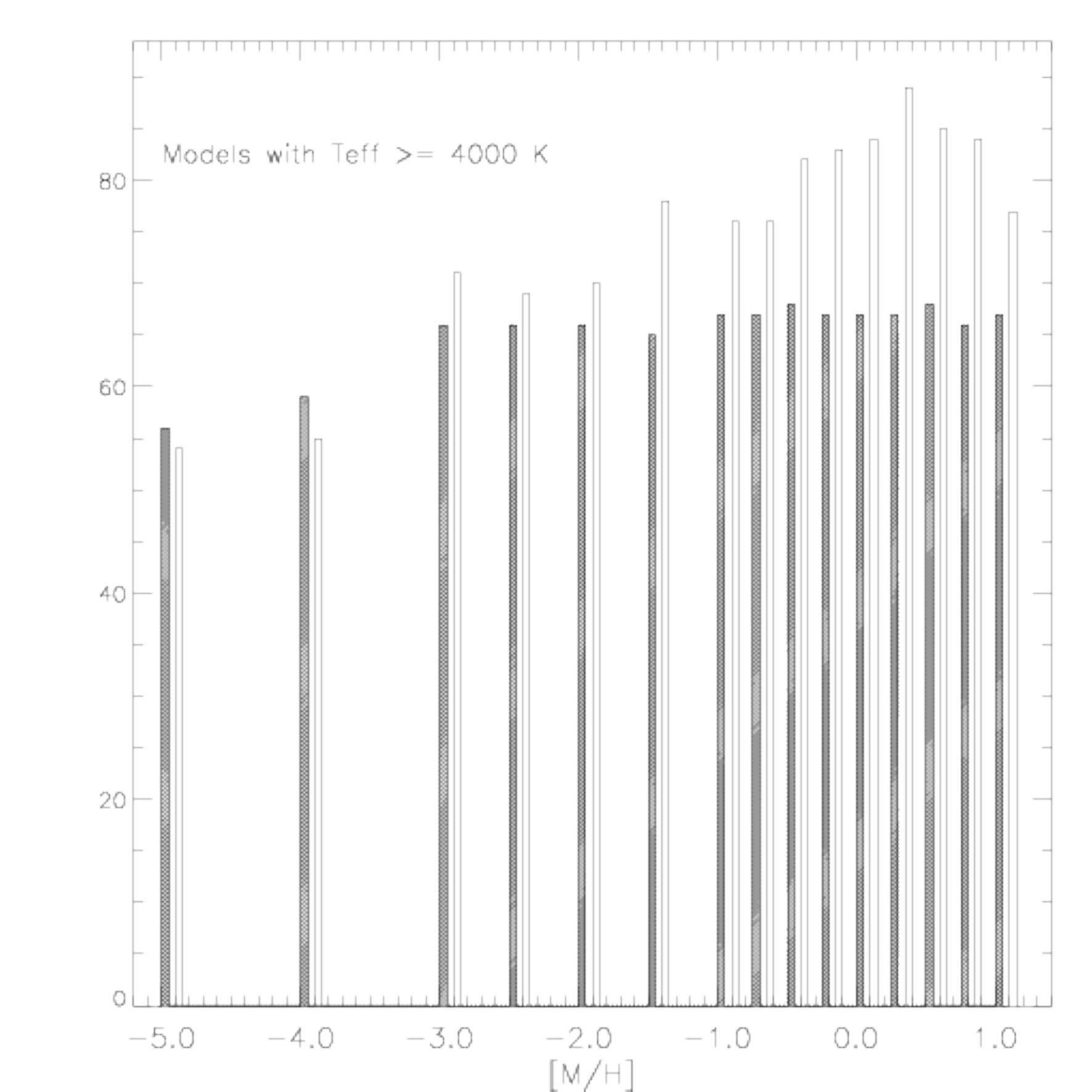}  
\caption{Distribution of the selected MARCS model atmospheres with respect to the metallicity for effective temperatures
cooler (left panel) or hotter (right panel) than 4\,000~K.
Filled and blank histograms refer to the adopted geometry for the radiative transfer, i.e. plane-parallel or spherical, respectively.}
\label{FigMeta}
\end{figure*}

Then, the selected models cover the mean metallicity (\meta \, hereafter, except the \alfa-elements) range
from -5.0 to +1.0, spanning 15 different values
(see Fig.~\ref{FigMeta} \& \ref{TabGrid}). 
We point out that, throughout this paper and the AMBRE project, the mean metallicity refers
to solar scaled abundances of all elements heavier than He.
It should also be noted that the adopted metallicity range allows for analysis of very different types of
Galactic stellar population, including extremely metal poor halo stars.

Finally, a selection has been performed on the enrichment in \alfa-elements. 
As in \cite{G08}, the chemical species O, Ne, Mg, Si,
S, Ar, Ca, and Ti are considered as \alfa-elements and we consider that they 
vary in lockstep.
We remind the reader that \cite{G08} have shown that the effects of an \alfa -enhancement
on the temperature structure in the atmosphere become significant below \T $<$ $\sim$4\,500~K for dwarfs and giants
(leading to larger temperature gradients in the deep layers and to higher surface heating, both being mainly
caused by TiO lines). For the present AMBRE grid, 
at a given metallicity, we only selected the MARCS
models with an \alfaFe \, enrichment consistent with the chemical
properties of most of the galactic stars, i.e. 
\alfaFe = 0.0 for \meta $\ge$ 0.0, 
\alfaFe = +0.4 for \meta $\le$ -1.0 and 
\alfaFe = -−0.4 $\times$ \meta for -−1.0 $\le$ \meta $\le$ 0.0.

Regarding the assumed geometry in these calculations (model atmospheres
and synthetic spectra), plane-parallel models have been
considered for +3.5 $\leq$ \g $\leq$ +5.5 (in these cases, the atmospheric extension 
is negligible relative to the radius, and thus sphericity effects
can be neglected). These models have
a microturbulent-velocity parameter set to 1.0~km/s.
A spherical geometry has been considered when \g $\leq$ 3.0 since sphericity effects may
be important for low masses and/or low \g \, (see
\cite{Heiter} and \cite{G08}). For these gravities, a mass of 1.0~M$_\odot$
and a microturbulence parameter of 2.0~km/s have been considered.
Fig.~\ref{FigMeta} presents the distribution of these selected model atmospheres with respect to 
their metallicity and the adopted plane-parallel/spherical geometry.

On the other hand, we emphasize that all the selected MARCS models have been computed assuming a
mixing length parameter $\alpha = l/H_p$ = 1.5 (\cite{G08}) and solar abundances
from Grevesse et al. (2007) that reflect recent 3D hydrodynamic, LTE-departures and improved atomic and molecular data
for the solar modelling.\\

\begin{figure*}[ht]
\includegraphics[width=19cm,height=19cm]{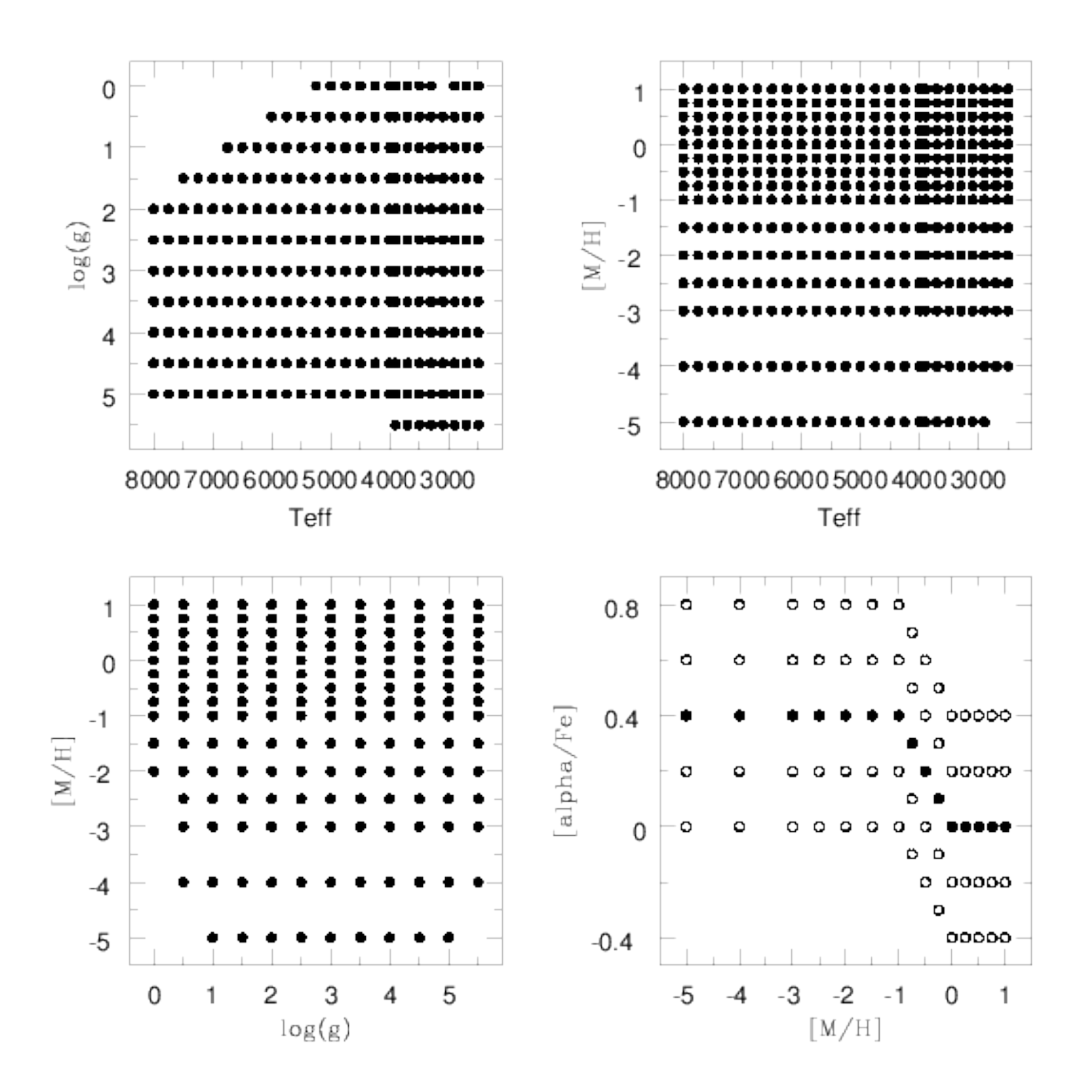} 
\caption{Distribution of the AMBRE synthetic spectra grid in the atmospheric parameters and \alfaFe \, space. We reiterate that only
one value of \alfaFe \, for every \meta \, was adopted during the model atmosphere selection process.
This is illustrated in the bottom right panel where the AMBRE spectra computed from MARCS models
with consistent \alfaFe \, ratios
are plotted with filled circles while open circles refer to all the other AMBRE spectra computed with
atmosphere models having inconsistent \alfaFe \, ratios.}
\label{FigGrid}
\end{figure*}

In total, 3\,358 MARCS models were selected (2\,115 models with \T$\geq$4000~K and 1\,243 models with \T$<$4000~K).
We point out that a few models are missing (all the combinations of
the atmospheric parameters were not available) due to the approach to the Eddington
flux limit and/or poor convergence of the MARCS code (see \cite{G08}).
The ranges of the atmospheric parameters of the AMBRE grid are summarised in Tab.~\ref{TabGrid} and illustrated in 
Fig.~\ref{FigGrid} (see also Fig.~\ref{FigMeta}).

\section{The AMBRE grid of synthetic spectra}

From the atmosphere models presented in the previous section, we computed
the final grid that consists of 16\,783 flux normalised and absolute-flux spectra
(10\,575 spectra have \T $\geq$ 4\,000~K and 6\,208 have \T $<$ 4\,000~K;
it has to be pointed out that 7 cool star spectra with available model atmospheres are missing due to numerical issues).
For reasons of consistency with the MARCS model atmospheres of \cite{G08}, 
the spectra have been computed assuming the solar chemical composition of Grevesse et al. (2007)
and similar geometry, mass and microturbulence parameter (see previous Section).

In the spectrum calculations, five different \alfaFe \, enrichments
were adopted for any selected MARCS model. The adopted \alfaFe \, values vary
from -0.4 to +0.4~dex (step of 0.2~dex) with respect to the canonical values that 
correspond to the original abundances of the selected MARCS models (see Tab.~\ref{TabGrid}
and Fig.~\ref{FigGrid}, bottom left panel).
Therefore, these computations were performed
   with chemical abundances that are sometimes inconsistent with
   those adopted in the calculation of the model atmosphere structure.
However, for each synthetic spectrum, all continuous and line opacities were recomputed from the model structure
with the right chemical composition 
(see Subsect.~\ref{SectAlfa} for an estimation of the errors induced by this assumption).

The delivered spectra cover the wavelength range between 3\,000 and 12\,000~\AA \, with a constant wavelength 
step of 0.01~\AA \, (900\,000 pixels in total). 
With the Nyquist-Shannon sampling condition, the adopted spectral sampling corresponds to an equivalent resolving power 
($R = \lambda / \Delta \lambda$)
larger than 150\,000 in the ultraviolet/blue spectral domain to $\sim$~600\,000 in the near infrared region.
We point out that we decided not to perform any convolution (spectral resolution and instrumental profile, 
stellar rotation, macroturbulence,...)
for these spectra in order to keep an easy and fast adaptation to the properties of any spectrograph
and/or astrophysical application.
The adopted spectral domain and sampling indeed cover the wavelength ranges
explored by all the ESO optical spectrographs, which are the main targets of the AMBRE project.
This spectra grid is also useful for almost any other astronomical optical spectrograph since 
spectra with any lower resolving power (or large-scale broadening) can be easily computed from the provided grid.
However, we warn the future users of this grid that the adopted wavelength step might not be well adapted for very high-spectral
resolution applications. Indeed, the narrowest lines could be slightly under-sampled in the present grid.
This has been tested by computing a few spectra with a wavelength step ten times smaller than the one of the AMBRE grid
(i.e. 0.001~\AA) and then by degrading these spectra to different spectral resolutions. Small differences between these spectra
having a better sampling and the AMBRE ones are found for spectral resolution larger than 50\,000 (they can reach
2-3\% in relative flux in the blue when $R = 100\,000$ or higher). 

All computations were performed in serial mode on the 952~cores (2.4Ghz Opteron) of the high-performance computing facility SIGAMM
of the Observatoire de la C\^ote d'Azur. The total CPU time used by this project was about 10$^5$~hours ($\sim$11 years if
one would have used a single CPU!). This large CPU time is almost completely caused by the 
computation of the opacities induced by the $\sim 10^7$ molecular lines in the different atmospheric layers, 
and it explains 
why these lines have not been considered in most of the synthetic spectra grids available so far.

Finally, the synthetic spectra are given in absolute surface flux at each wavelength bin (in units of erg/cm$^2$/s$^1$/\AA)
and in normalised flux, relative to the local continuum flux (no units). The continuum-normalised spectrum 
is derived by dividing the absolute flux spectrum by the computed continuum flux in every spectral bin.
The size of one spectrum file in FITS format is 3.5~Mbytes (about 10 times more in ASCII format) and the total grid of 16783 spectra has a size of 59~Gbytes in FITS.

\section{Contribution of the different chemical species to the spectrum flux}

\begin{figure*}[ht]
\includegraphics[width=9.2cm]{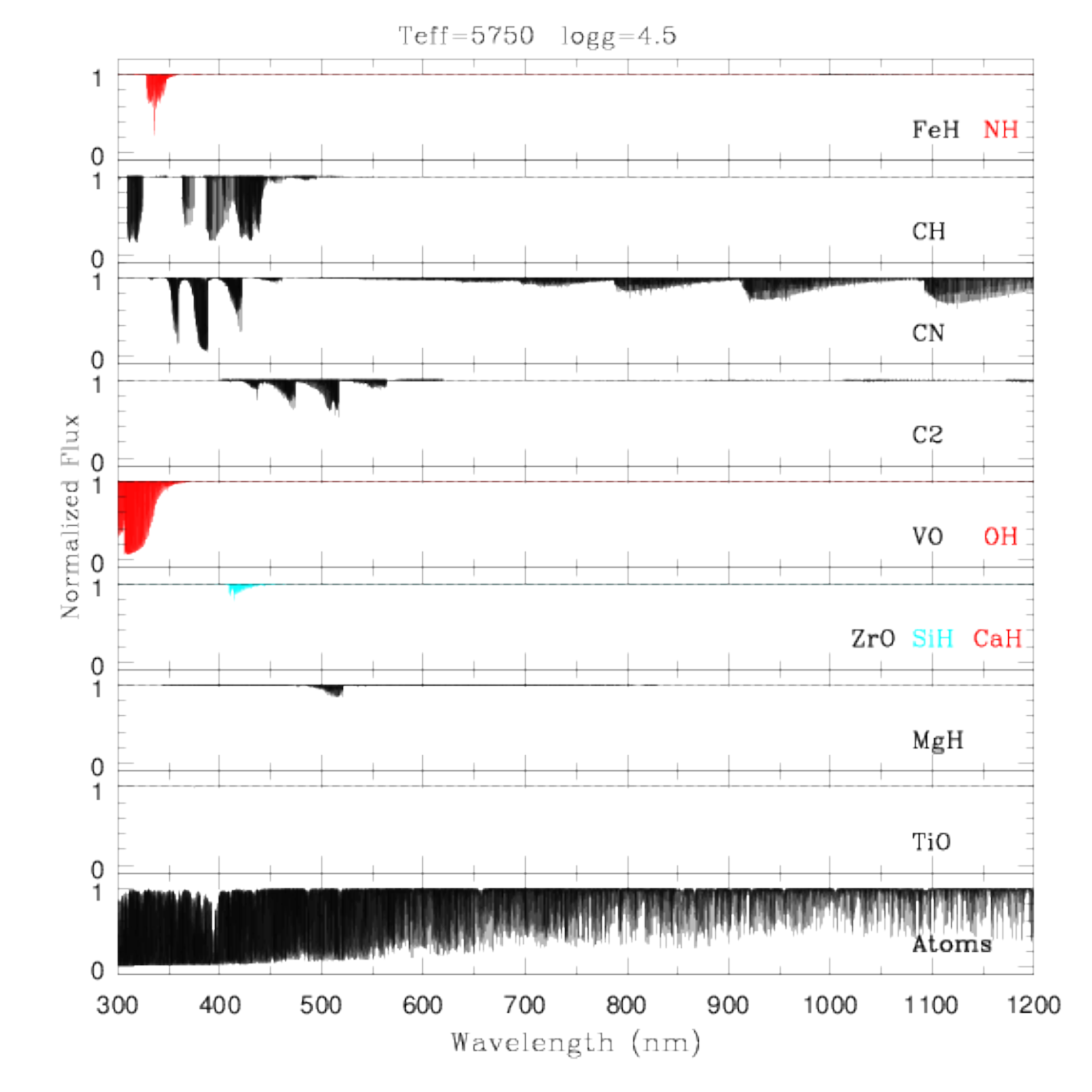} 
\includegraphics[width=9.2cm]{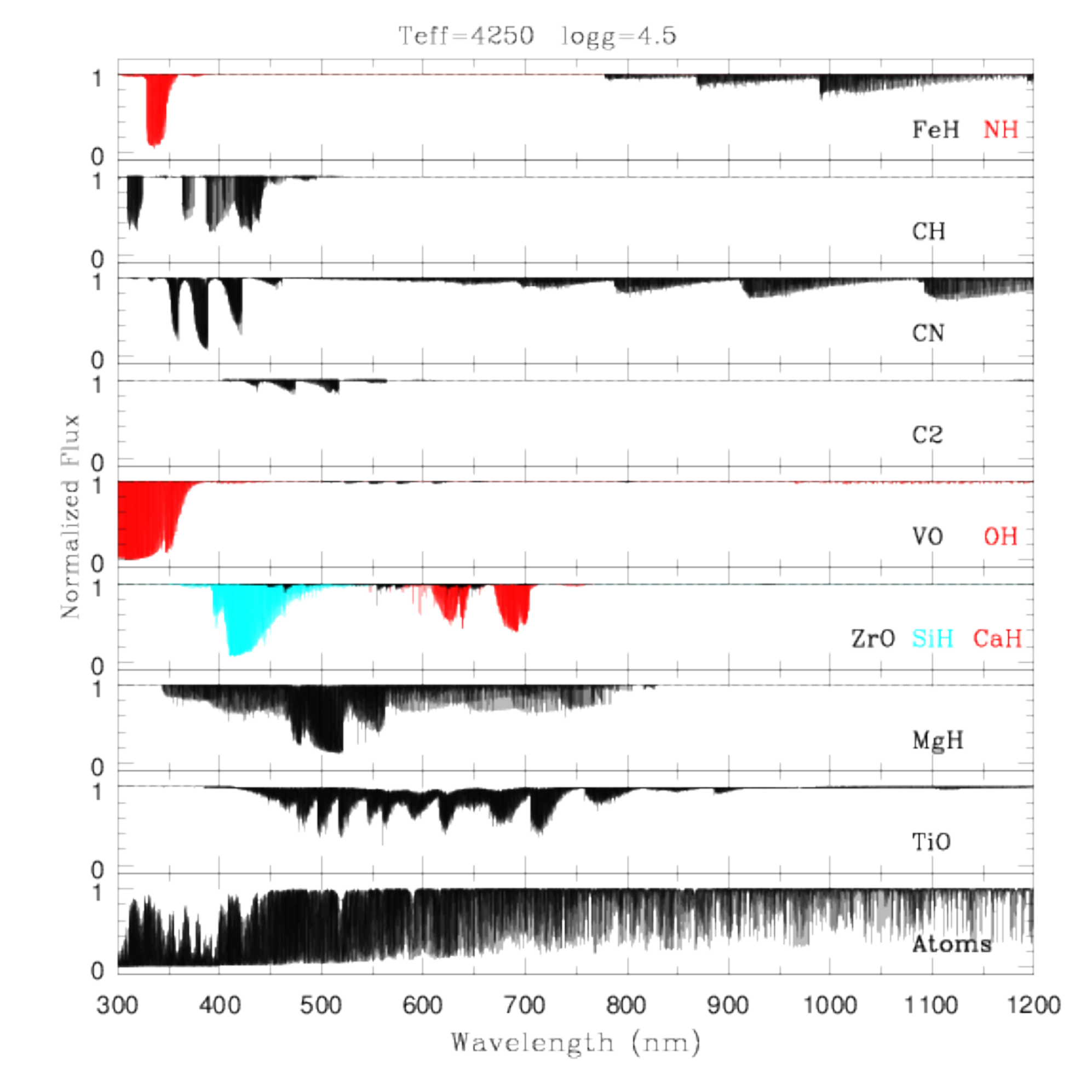} 
\includegraphics[width=9.2cm]{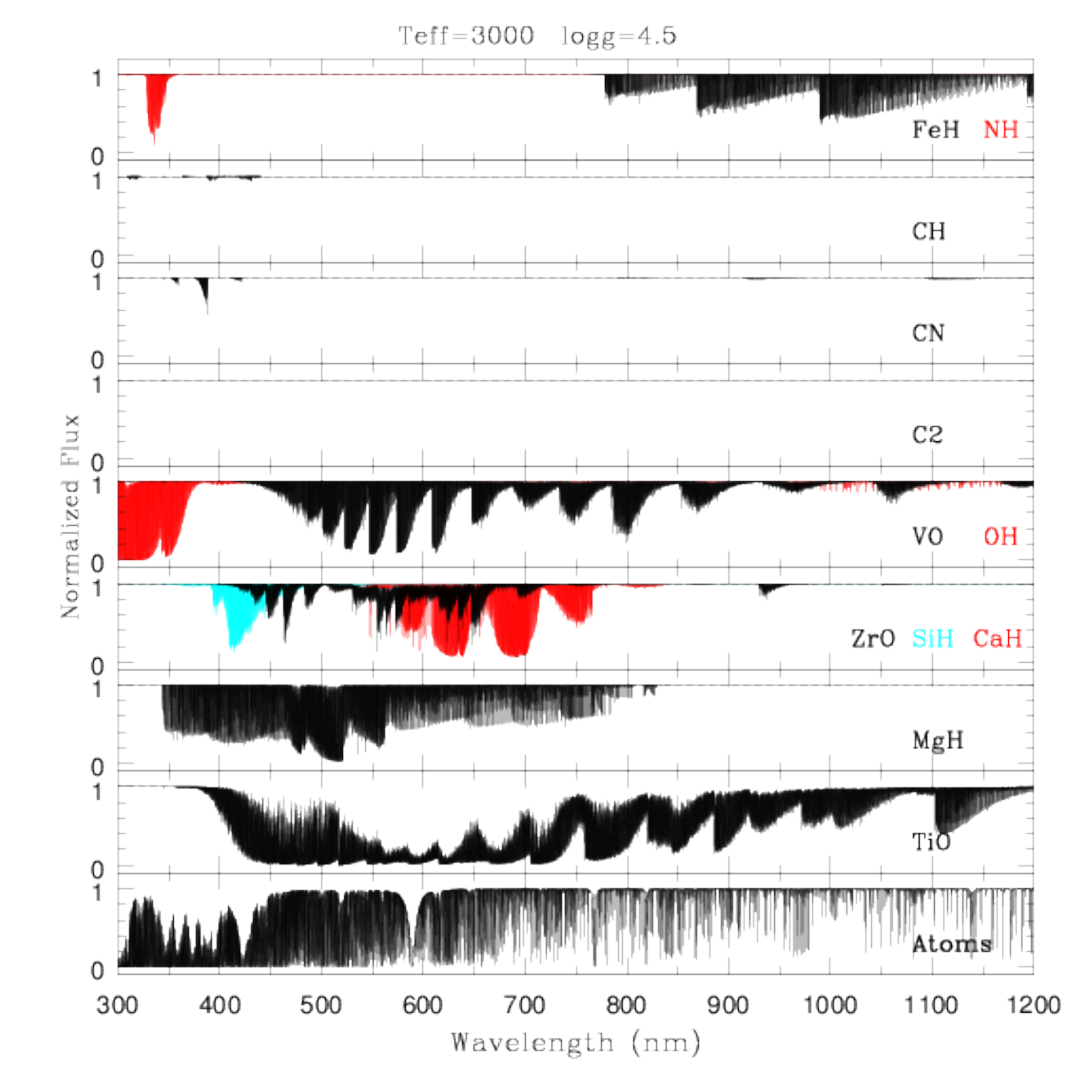}
\includegraphics[width=9.2cm]{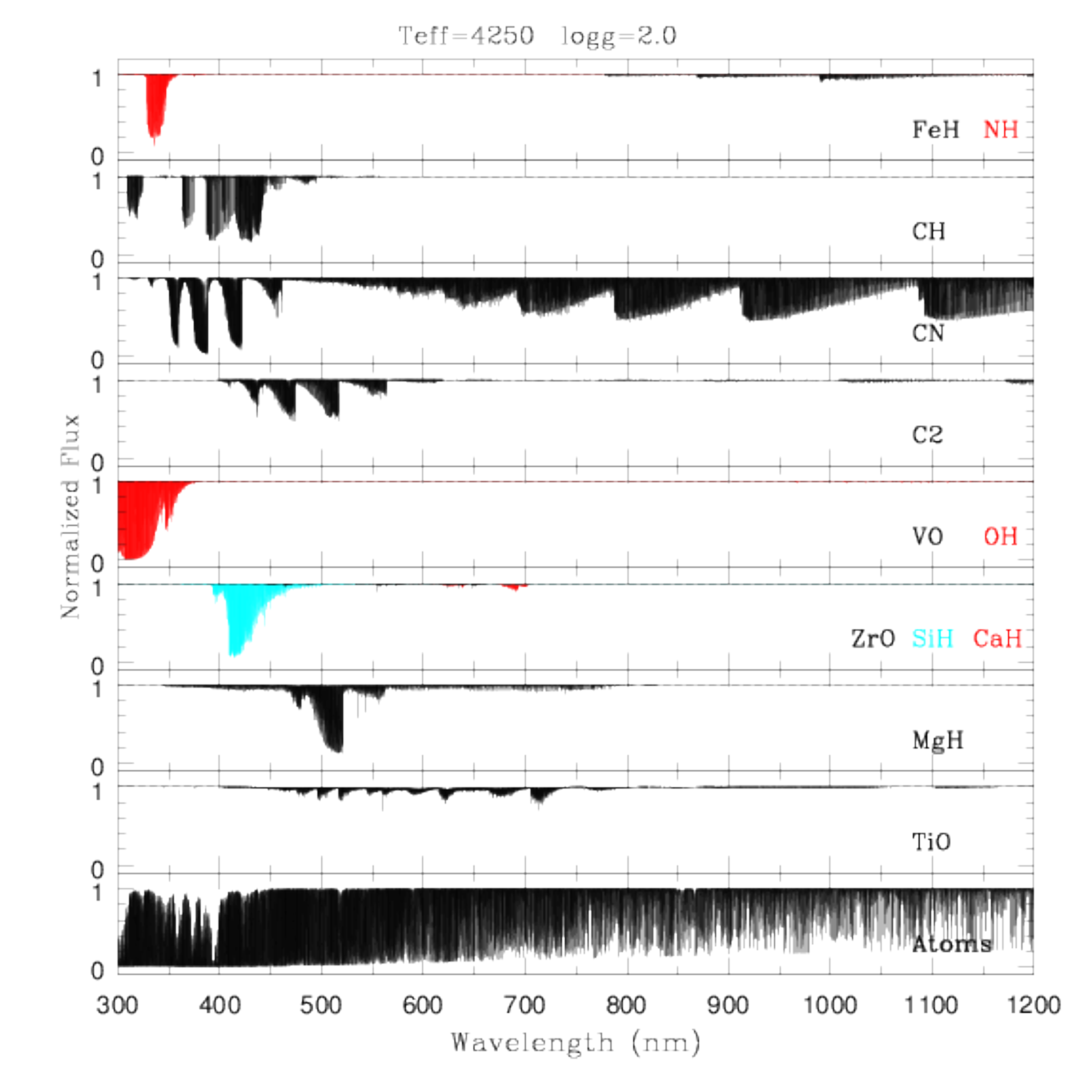}
\caption{Normalized spectra computed with only one line specy (all atoms or individual
 molecules) included in the synthesis. This figure illustrates the line blocking by different species
for metal-rich stars with solar \alfa -enrichment (\meta = +0.5 \& \alfaFe = 0.0). Examples are shown for a solar type star (\T = 5750~K \& \g = 4.5; top left panel), a cool dwarf star (\T = 4250~K \& \g = 4.5; top right panel), a very cool dwarf star (\T = 3000~K \& \g = 4.5; bottom left panel) and a cool giant star (\T = 4250~K, \g = 2.0; 
bottom right panel). To ease the comparison, molecules are grouped with respect to their carbon-rich (C$_{2}$, CN and CH) or \alfa -rich 
(TiO, ZrO, MgH, CaH, SiH and OH) nature.}
\label{FigContribution}
\end{figure*}

In order to study the contribution of the different chemical species 
to the total opacity, we computed test spectra by considering separately the different
linelists. 
This analysis has been performed on the spectra of
different types of stars that are slightly enriched in metals (\meta = +0.5) in order
to enhance the different contributions. We show in Fig.~\ref{FigContribution} the 
line blocking by different species (atoms and molecules).
The atomic lines are the dominant source of line opacity for dwarf and giant stars hotter than $\sim$4\,000~K.
Molecules become the strongest contributors for cooler stars.
More precisely, carbon-based diatomic molecules (C$_{2}$, CN and CH) have an important contribution with respect
to other molecules for \T $>$~4\,500~K. 
However, at lower temperatures for dwarf stars, the molecular contribution is then dominated by molecules composed
of \alfa -elements (TiO, ZrO, VO, MgH, CaH, SiH and OH). Their blanketing effect thus strongly increases
at such low effective temperatures. These molecules (and TiO in particular) may therefore strongly affect
the temperature structure in the atmosphere.
On the contrary, we notice that cool giant spectra are more dominated by carbon-rich
molecules than their dwarf counterparts with similar effective temperatures.\\

\begin{figure*}[ht]
\includegraphics[width=9.2cm]{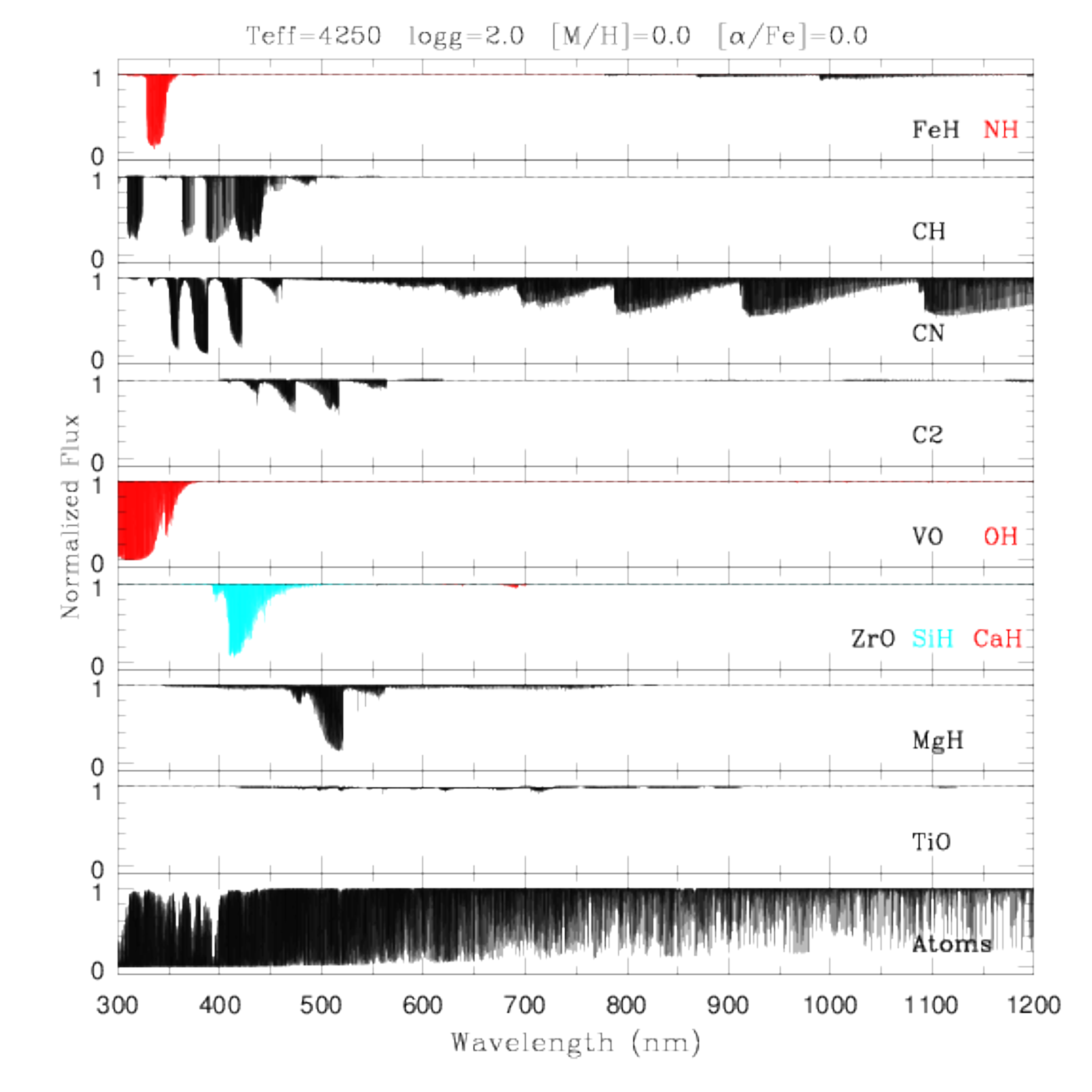} 
\includegraphics[width=9.2cm]{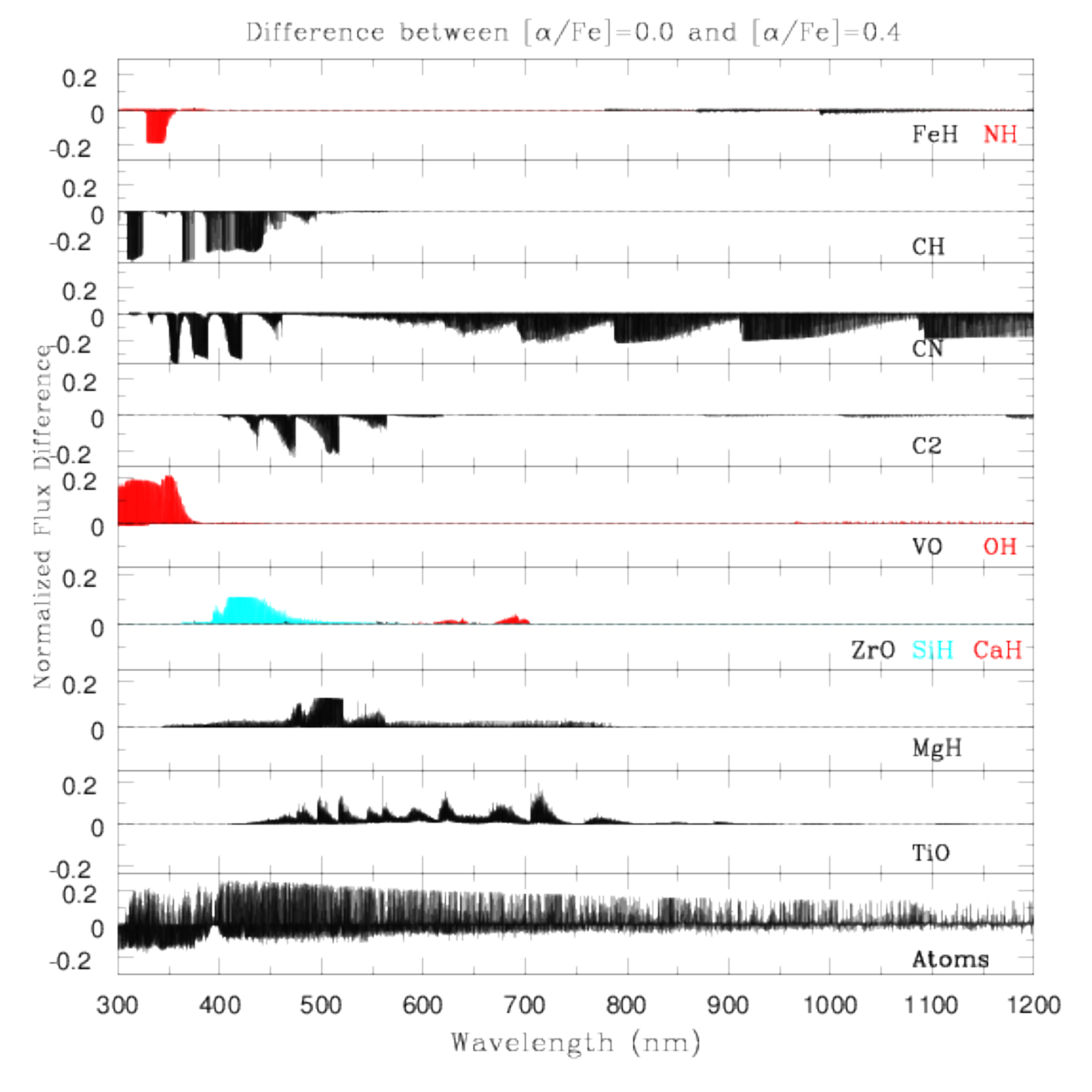}
\caption{Spectrum comparison of typical metal-rich giant stars observed towards the galactic bulge
with two different \alfa-enhancements. The main atmospheric parameters are \T = 4250~K, \g = 2.0 and \meta = 0.0.
Left panel: Same as Fig.~\ref{FigContribution} for normalised spectra computed with only one line specy in the synthesis
and assuming \alfaFe =+0.0 as detected
in some bulge directions and in the galactic thin disk.
Right panel: Line blocking differences between the left panel case (\alfaFe =+0.0) and a bulge component star having \alfaFe =+0.4. It can be seen that increasing the abundance of \alfa -elements leads to weaker lines of non
\alfa -atoms and molecules since the continuum flux becomes lower (see Sect.~4 for more details).}
\label{FigContributionBulbe}
\end{figure*}

\begin{figure*}[ht]
\includegraphics[width=9.2cm]{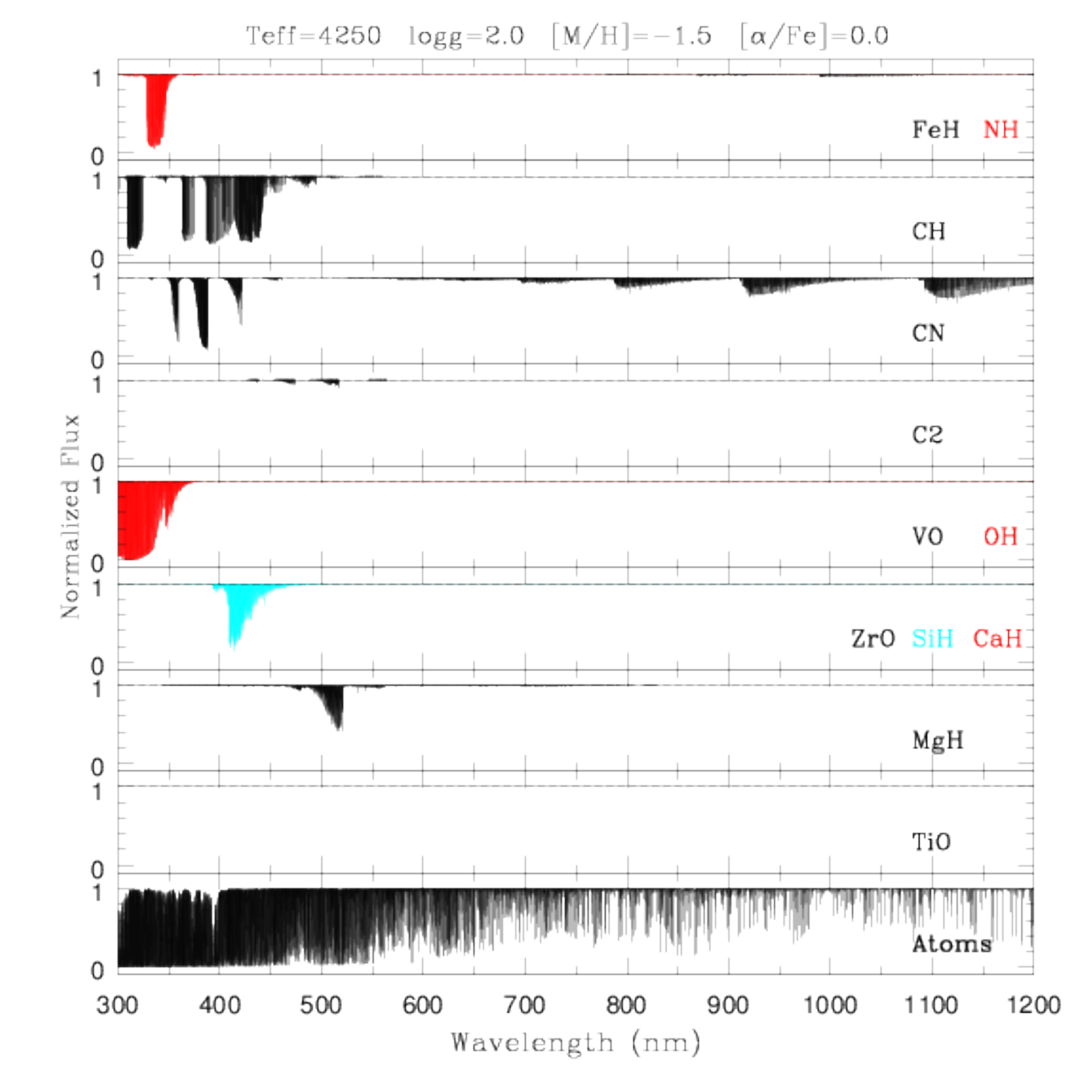} 
\includegraphics[width=9.2cm]{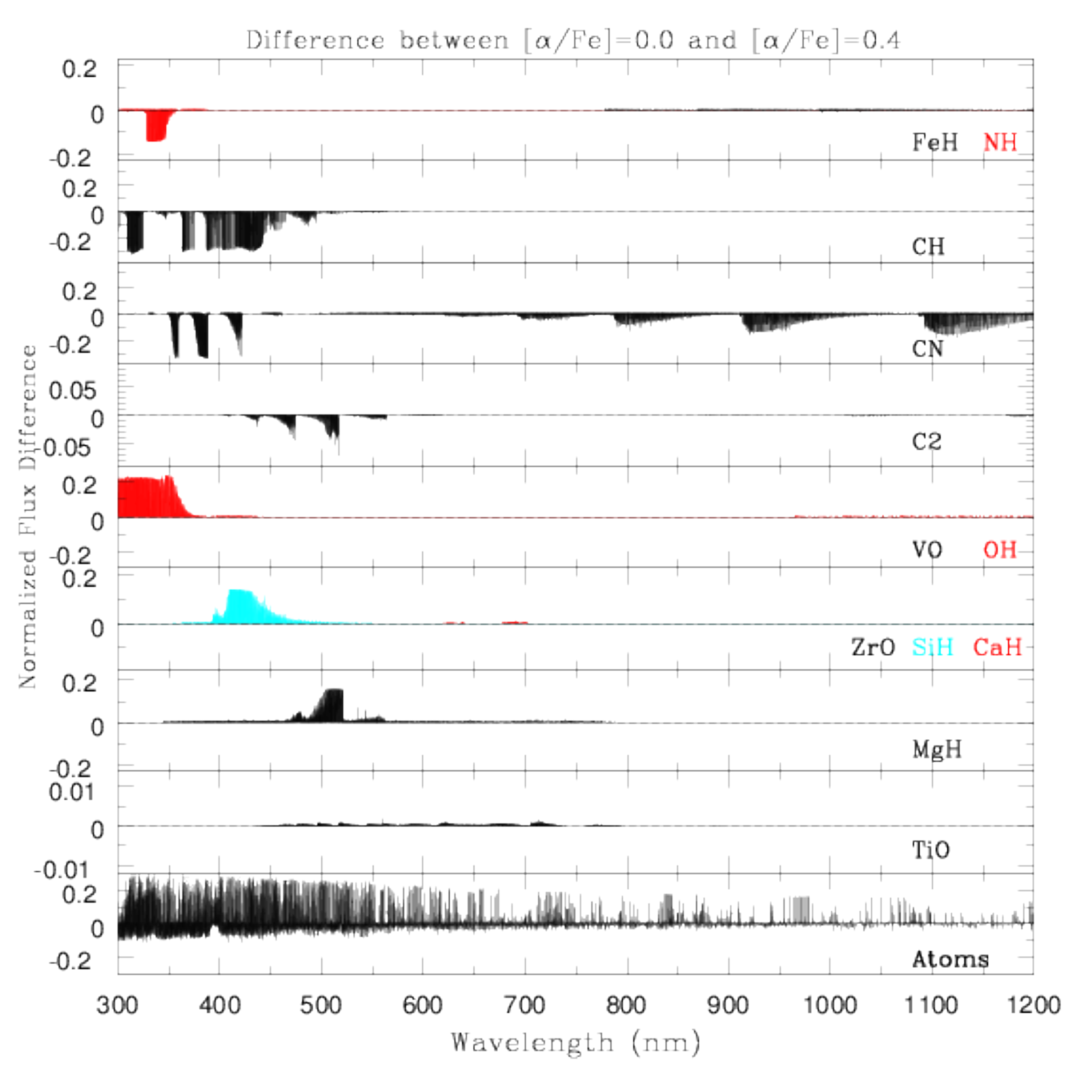}
\caption{Same as Fig.~\ref{FigContributionBulbe} for typical metal-poor giant stars of the galactic halo with two different \alfa-enhancements. The main atmospheric parameters are \T = 4250~K, \g = 2.0 and \meta = -1.5.
Left panel: normalised spectra computed with only one line specy in the synthesis and assuming \alfaFe =+0.0 as detected in halo giants probably originating from accreted dwarf galaxy satellites.
Right panel: Line blocking differences between the left panel case (\alfaFe =+0.0) and a giant having \alfaFe =+0.4
as measured in halo stars formed in-situ (see caption of Fig.~\ref{FigContributionBulbe}, Right panel). }
\label{FigContributionHalo}
\end{figure*}

We have also examined the line blocking of atomic and
molecular lines for two typical giant stars of the galactic bulge and halo (Fig.~\ref{FigContributionBulbe}
\& \ref{FigContributionHalo}, respectively). In these galactic components, one can encounter
stars with slightly different \alfaFe -enrichments. For that purpose, we have computed spectra with similar atmospheric
parameters (\T, \g, \meta) and different \alfa-enrichments. We point out that in all these comparisons the considered model 
atmosphere has an \alfaFe \, ratio that is consistent with the one
adopted in the synthetic spectrum calculations (this is not the case for every AMBRE spectra, see Sect.~\ref{SectAlfa}
for a discussion of this assumption). As we have mentioned already in Sect~2.2, the atmospheric temperature structure 
of cool dwarfs and giants is strongly affected when \alfa -enhancement are considered (see \cite{G08}).

For instance, Fig.~\ref{FigContributionBulbe} compares
the spectra of a metal-rich (\meta = 0.0) giant of the bulge with \alfaFe =+0.4 and of a typical disk giant 
(\alfaFe = +0.0) that could be detected in the same line-of-sight. Such variations in \alfa-enrichments can also be encountered
in bulge giants observed at different galactic latitudes (Gonzalez et al., 2011). 
It can be seen that differences as large as 10-20\% in relative flux are found for the atomic lines and most molecular lines.
These differences affect almost any spectral domain, even those dominated by non-\alfa -rich molecules. 
Furthermore, metallic lines (except those of \alfa -elements) and FeH, NH or carbon-rich molecular lines 
appear stronger in the \alfaFe = +0.0 spectrum.
In other words, several lines of non \alfa \, elements appear weaker in the \alfa -enhanced spectra,
although their adopted abundance is kept constant. This is caused by the fact that \alfa -elements 
(and, particularly, Mg, Si and Ca) are important sources
of electrons. They therefore strongly contribute to the total continuous opacity,
dominated by the H$^-$ ion in the studied stars (with respect to HI Rayleigh scattering). 
Thus, the continuum flux level of \alfa -enhanced spectra is lower. 
For instance, in the two spectra of Fig.~\ref{FigContributionBulbe}, the
continuum flux is 25\% higher in the blue for the \alfaFe =+0.0 spectrum with respect to the \alfaFe =+0.4 one (this difference
decreases towards longer wavelengths and is less than 2\% in the near infrared). 
In summary, increasing the abundance of \alfa -elements therefore leads to weaker lines of non \alfa -atoms and molecules
in normalised spectra.
 
Similar effects (caused by identical reasons)
can be seen in Fig.\ref{FigContributionHalo} where the differences between two halo giant spectra (\meta = -1.5) are shown: one formed in-situ
(\alfaFe = +0.4) and one possibly accreted from a satellite dwarf galaxy (\alfaFe =+0.0; as found for instance in Fornax or Sculptor, see Tolstoy et al., 2009). Again, it can be interestingly noted that the main differences between both spectra are found for the atomic
lines and those of NH, CH, CN and C$_2$ (their
lines appear deeper in the \alfaFe = +0.0 spectrum) whereas the OH, SiH and MgH lines appear much stronger for \alfaFe = +0.4.

As a consequence of all these discussions, it should be pointed out again that the integrated spectrum
of galactic stellar components or stellar populations found in external galaxies
containing large proportions of cool main-sequence and RGB stars
is thus affected by their respective enrichment
in \alfa -elements.
This should be carefully considered when interpreting such integrated spectra.

\section{Inconsistent \alfaFe \, ratios between the model atmospheres and the 
synthetic spectra}
\label{SectAlfa}

We recall that several spectra of the AMBRE grid have been computed from model atmospheres
with inconsistent enrichments in \alfa -elements with respect to the ones adopted in the
spectral synthesis (a difference, noted $\Delta$\alfaFe \, hereafter, of $\pm$0.2~dex
or $\pm$0.4~dex can be encountered)\footnote{The tests performed in
the previous section do not suffer from this inconsistency.}. This shortcoming was adopted in order to compute a spectra grid
with several variations in \alfaFe \, (even if no MARCS models were available) 
in order to be able, for instance,  to automatically derive the \alfaFe \,
content of the analysed spectra as required by several galactic archaeology projects. In 2009, when we recovered the MARCS models, too few models
with several \alfaFe -values for a given metallicity were available (only some \alfa -poor models
were existing at that time). The situation is now improving
since more \alfa -enhanced or \alfa -poor models have been made available since summer 2010
and many 
more MARCS models should be computed in the near future (as they are required by the new galactic surveys
as GES or Gaia). When such MARCS models will be available, the adopted assumption could
be easily given up when computing a new grid.\\

\begin{figure*}[ht]
\includegraphics[width=9.2cm]{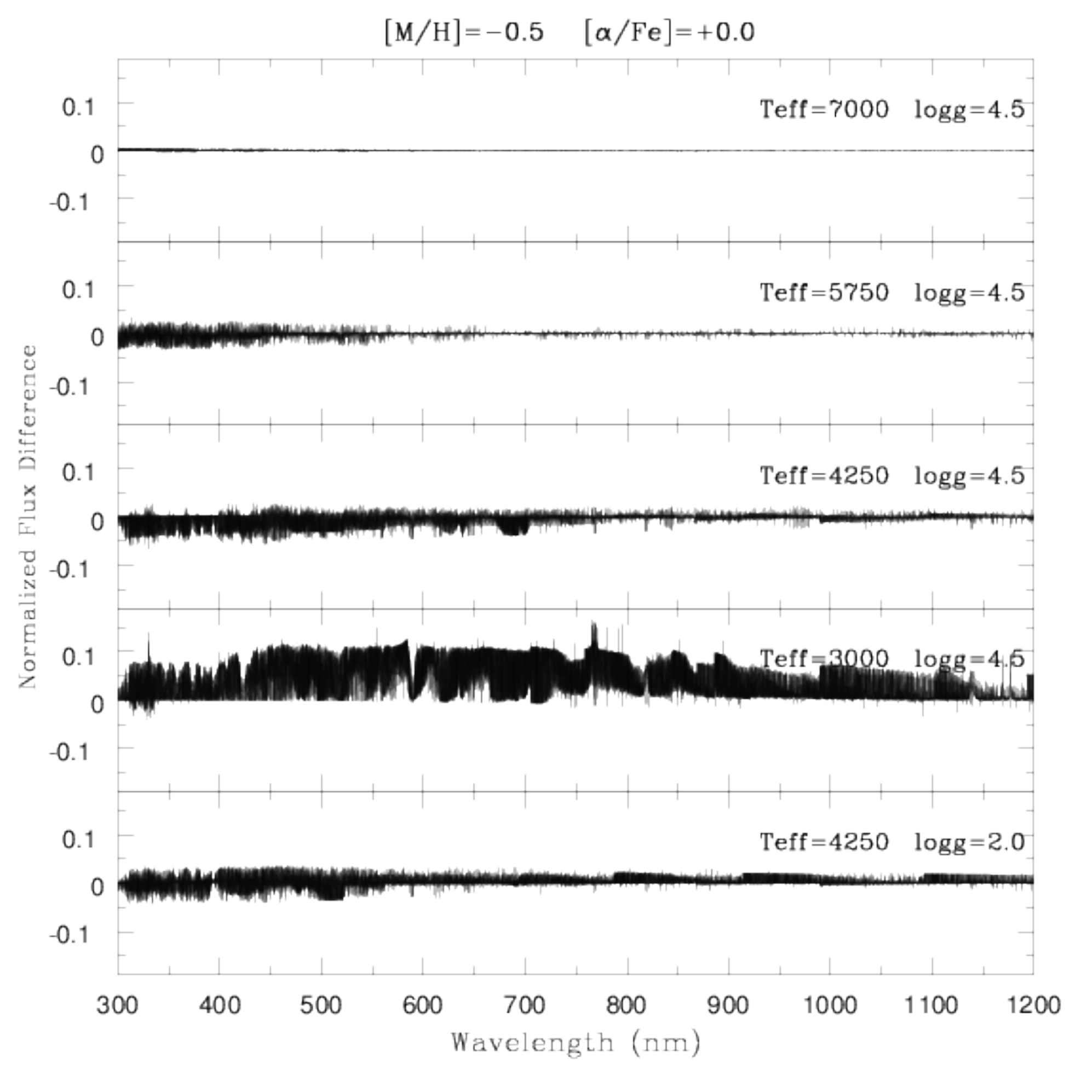}
\includegraphics[width=9.2cm]{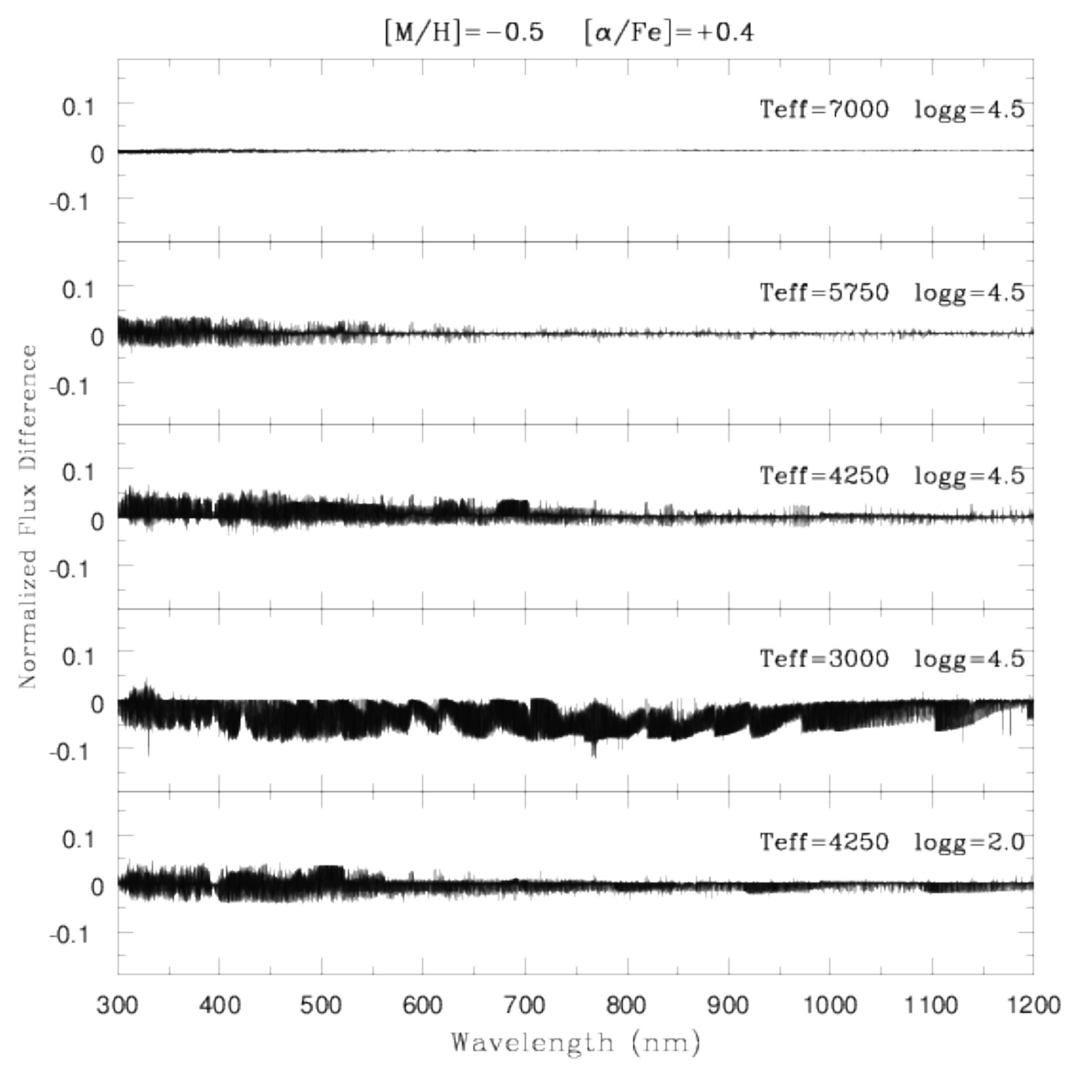}
\caption{Differences between the {\it exact} spectra (computed by adopting consistent \alfaFe \, ratios in the model atmosphere
and the spectral synthesis) and the corresponding AMBRE ones. The computed differences correspond to the {\it exact} spectrum minus the corresponding AMBRE spectrum.
The \alfaFe \, ratio of the adopted model atmosphere for the AMBRE spectra of the left panel is always +0.2~dex
while the spectral synthesis have been computed assuming \alfaFe = +0.0.
In the right panel, the adopted model atmosphere for the AMBRE spectra have always \alfaFe = +0.2
while we assumed \alfaFe = +0.4 for the spectral synthesis.
The effective temperature and surface gravity of the {\it exact} and AMBRE spectra
is indicated in the different panels. All spectra ({\it exact} and AMBRE ones)
have \meta = -0.5 and \alfaFe = +0.0 or +0.4 as indicated in the figure
title (left and right panels, respectively).}
\label{FigCompar1}
\end{figure*}

Therefore, the AMBRE grid contains spectra for which the continuous and line opacities
were computed from $(T, \tau)$-laws with different \alfaFe \, values.
Since most of the \alfa -elements (as Mg, Si and Ca) are important electron-donors,
their individual abundances directly impact the amount of H$^-$ ions (a major
opacity source in cool stars) formed in the different atmospheric layers
and, therefore, the computed atmospheric structure.
Morover, another effect will result from this assumption since the TiO molecule (composed
by two \alfa \, atoms) strongly
affects the atmospheric structure of the coolest models (\T $< \sim$ 4\,500~K) as already mentionned in the previous section.
In order to estimate the consequences of this assumption, 
we have computed the differences between some AMBRE spectra and others computed without
adopting this assumption, i.e. with consistent
\alfaFe \, ratios in the model atmosphere and in the spectrum computation (called {\it exact} spectra, hereafter).
We show in Fig.~\ref{FigCompar1} different examples in which the AMBRE spectra
have been computed with model atmospheres having an \alfaFe \, differing
by $\Delta$\alfa =$\pm$0.2~dex with respect to the {\it exact} spectra. It can be seen that the agreement is very good (differences are lower than a few percents in normalised flux at any wavelength) for \T $>$ 4\,000~K, whatever the value of the gravity is. The largest differences are found 
below $\lambda < 5\,000~\AA$ where most of the lines are encountered.
However, differences as large as $\sim$10\% over the whole optical domain 
appear for much cooler effective temperatures where many more lines contribute to the spectra.
We have also checked that these differences, which are probably produced by a 
continuum level not perfectly estimated in the tested AMBRE spectra,
 do not strongly vary when the stellar metallicity increases. Of course, for metal-poor
stars the agreement becomes much better.

\begin{figure*}[ht]
\includegraphics[width=9.2cm]{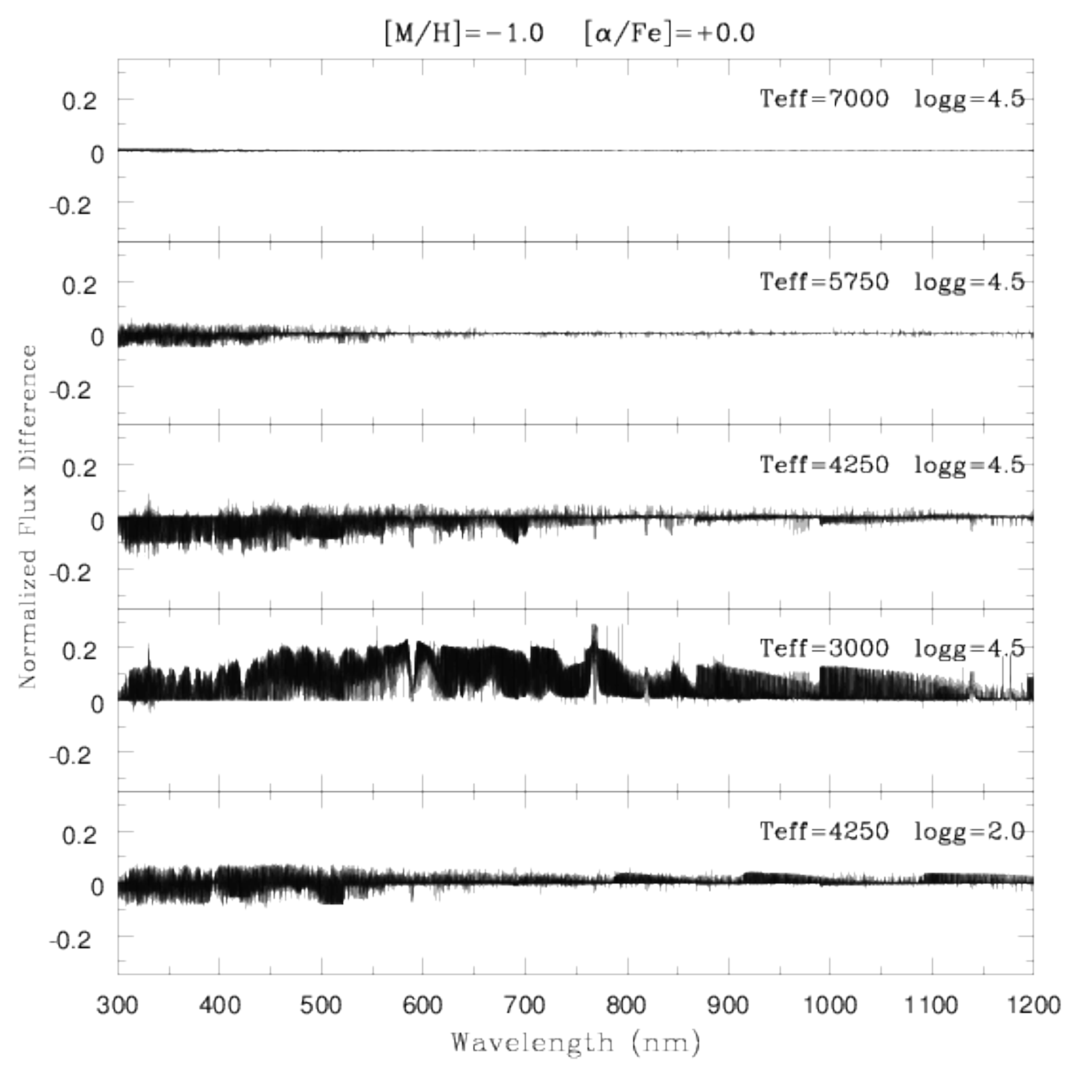}
\includegraphics[width=9.2cm]{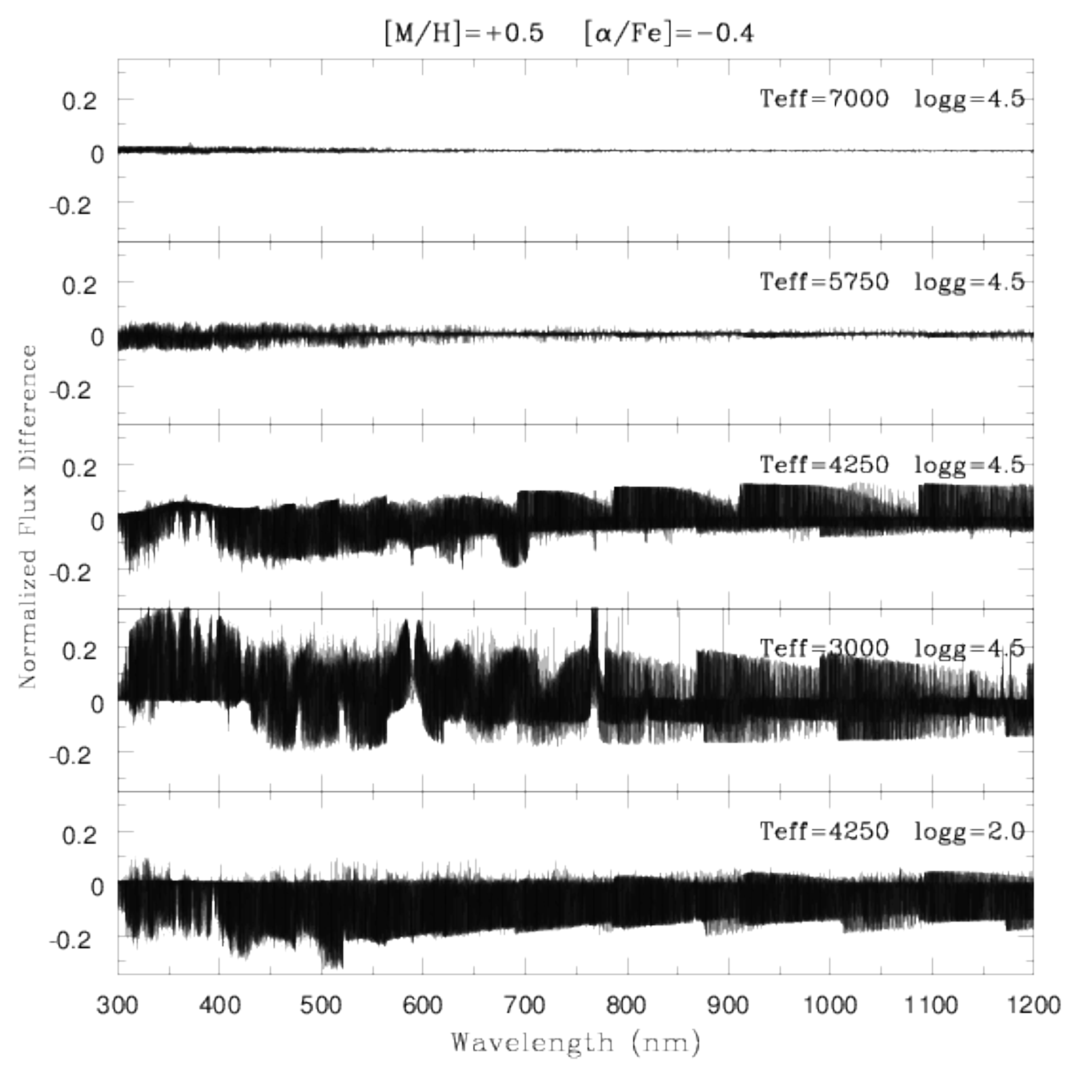}
\caption{Same as Fig.~\ref{FigCompar1} for \alfaFe \, of the AMBRE spectra differing by $\pm$0.4~dex
between the model atmosphere and the spectral synthesis. 
Left panel: typical metal-poor stars with low \alfa -enrichment. The \alfaFe \, ratio of the adopted model atmosphere for the AMBRE spectra is always +0.4~dex
while the spectral synthesis have been computed assuming \alfaFe = +0.0.
Right panel: very metal-rich stars with extremely low \alfa -enrichment. 
The adopted model atmosphere for the AMBRE spectra have always \alfaFe = +0.0
while we assumed \alfaFe = -0.4 for the spectral synthesis.}
\label{FigCompar2}
\end{figure*}

The effects of larger differences in \alfaFe -ratios between the adopted atmospheric
structure and the spectral synthesis are shown in Fig.~\ref{FigCompar2}
for a rather metal-poor star and a very metal-rich one (for which the studied effects
are extreme).
For rather hot stars (\T $>$ 4500~K), the spectrum differences are still quite small
whatever the luminosity and the metallicity are. The differences are always lower
than a few percents for $\lambda > 5\,000~\AA$ and stay below $\sim$10\% more in the blue.
However, larger differences (up to 10-15\%) are encountered for cooler stars
with effective temperature below $\sim$4\,000~K. For even cooler stars (M spectral types),
the quality of the AMBRE spectra with $\Delta$\alfaFe =$\pm$0.4~dex between the model
atmospheres and the spectral synthesis is rather poor and they should be used very
cautiously (or avoided when possible).

Furthermore, in order to better quantify the differences caused by a different adopted \alfaFe \,
ratio between the model
atmospheres and the spectral synthesis, we have computed the distance (by a $\chi ^2$ estimation)
between
an {\it exact} spectrum and several AMBRE ones with close stellar parameters.
Practically, we considered AMBRE spectra having \T , \g, \meta, and \alfaFe \, values between 
$\pm$1\,000~K, $\pm$0.5~dex, $\pm$0.5~dex and $\pm$0.4dex, respectively,
around the parameters of the {\it exact} spectrum. We thus computed distances between several hundred
of AMBRE spectra and a given {\it exact} one. All the stellar parameter combinations of
Fig.~\ref{FigCompar1} \& \ref{FigCompar2} were considered.
Such a test allows a check as to whether an algorithm could automatically derive the right 
atmospheric parameters of an input spectrum (an {\it exact} one in this case)
using the AMBRE grid. 
It has been found that the right solution is always recovered
if the effective temperature is larger than $\sim$4\,500~K, whatever the \alfaFe \,
enrichment.
For cooler stars, when $\Delta$\alfaFe=$\pm$0.2~dex and \T $< 4\,000$~K, the right solution is also
perfectly recovered for dwarf stars, 
while giants are found $\sim$ 250~K hotter (one grid step in \T , all the other parameters
being well retrieved) when 
$\Delta$\alfaFe = -0.2~dex or with a surface gravity slightly larger (one grid step in \g , all the other parameters
being well retrieved) when $\Delta$\alfaFe = +0.2~dex.
For the coolest cases (\T $<$ $\sim$3\,700~K for dwarfs and giants with $\Delta$\alfaFe = $\pm$0.2~dex
or when  $\Delta$\alfaFe = $\pm$0.4~dex and \T $<$ 4\,250~K),
the closest spectra are always found within
one step grid in each atmospheric parameters. 
However, depending on the test cases, any of the four atmospheric
parameter can be erroneously derived although only \T \, or \g \, are affected
in most cases. In the worst cases (very cool stars and $\Delta$\alfaFe = $\pm$0.4~dex), 
two of these parameters can be simultaneously badly recovered
although always within one step grid.
We therefore conclude that, even in these extreme cases (very cool stars and/or large $\Delta$\alfaFe) where the accuracy of the parameter
determination has decreased, the main stellar properties are still enough well recovered in order to classify the observed stars.
It has also been found that the situation improved if only the reddest
part of the spectra ($\lambda > \sim$5\,500~\AA) is considered since a large amount of the atomic
and molecular lines are found at shorter wavelengths.

Finally, we would like to point out that
similar warnings 
have to be considered for absolute flux spectra.
However, most of these differences
would be partly or completely hidden for observed spectra with moderate or low Signal-to-Noise ratios ($\sim$~50 or lower)
and/or very metal-poor stars.

\section{Data access}
One of our goals when computing this grid was to offer it to the whole astronomical community.
It can therefore be easily recovered upon request to the authors (laverny@oca.eu). Each spectrum (relative
and absolute fluxes) can be obtained individually in FITS format (3.6Mb each).
The whole grid (16\, 783 spectra) has a total size of $\sim$60Gb.

Also a large part of this grid has been made publicly available through the POLLUX database 
of synthetic stellar spectra (\cite{Palacios}). Every AMBRE spectrum with 5\,000 $\le$ \T $\le$ 8\,000~K
can presently be recovered from the POLLUX website (http://pollux.graal.univ-montp2.fr/)
that contains 7\,940 AMBRE spectra, i.e.
about 50\% of the whole grid.

\section{Conclusion}
\label{conclu}

We have computed, and made publicly available, a very large
grid of FGKM star synthetic spectra covering the whole optical domain.
It includes almost any possible combination of the atmospheric parameters
\T  , \g , \meta \, and \alfaFe \, (including extreme
metal-poor and metal-rich stars). These spectra
are based on (and are mostly consistent with) the MARCS model atmospheres and include the most
complete atomic and molecular linelists available so far. 
We have shown that about 90\% of the provided spectra, including those spectra with extreme \alfaFe \, values
that have been computed with \alfaFe \, abundances inconsistent with those of the model atmospheres,
can be safely adopted for any use under these criteria: (i) the considered \alfaFe \, ratios differ
at maximum by $\pm$0.2~dex between the model atmosphere and the spectral synthesis, (ii) high \T \, ($\geq \sim$4\,250~K) and larger $\Delta$\alfaFe \, values, or (iii) the most metal-poor stars for any value of $\Delta$\alfaFe .

However, when $\Delta$\alfaFe=$\pm$0.4~dex and \T \, $<$ $\sim$4\,250~K, the spectra
should be considered very cautiously (or avoided when possible) although 
they are still similar enough to the {\it exact} ones (within one step grid in every parameter dimension)
to allow spectral classification.

We are well aware that such `classical' model atmospheres and synthetic spectra 
can be questioned when they are carefully compared to real stars, since the effects of the adopted different 
assumptions  (LTE, 1D, static atmospheric structure, varied abundances of individual elements,...) 
could be large for some specific cases.
The accuracy of the adopted linedata together with their systematics should
also be questionned.
For instance, for the coolest spectra of this grid (below $\sim$ 2\,800~K), 
we did not consider
model atmospheres and opacities caused by the formation of dust grains. Such an assumption
may at least affect the red part of the spectra and/or metal-rich models (see \cite{Chabrier}).
However, several of these effects could be partly corrected by adopting specific calibrations
defined with rather well known reference stars (see, for instance, the procedure adopted
within the AMBRE project; \cite{Worley}). Furthermore, more advanced models (including NLTE effects, 3D, hydrodynamics,...) are very far
from covering the whole range of stellar parameters considered in this work, and there exists no
other alternative than computing such `classical' grids.

However the provided grid consists of high resolution spectra.
They can be very easily adapted to any real observed spectra: either 
by slicing them over the covered wavelength ranges,
by convolution with a well adapted kernel and/or by adopting a specific sampling law
(see the application to the analysis of FEROS spectra in \cite{Worley}).
We have also preferred not to consider any stellar rotational
velocities and/or macroturbulence profiles in order to ensure
a very broad use.
This grid is at the basis of the whole
AMBRE project that consists of the analysis of most of the ESO
high-resolution archive spectra. All of the training phases for the
parametrisation algorithms have indeed been performed with these AMBRE spectra, 
adapted in turn to each of the four selected spectrographs including their different instrument set-ups.
This grid has also been adopted for several
tests of the classification algorithms developed for the Gaia/RVS spectra and for
the preparation of the Gaia-ESO Survey (selection of the GIRAFFE setups,
estimates of the uncertainties in the derived atmospheric parameters
and abundances,...).
Finally, other applications could easily be considered as, for instance,
the study of the integrated spectrum of the different Galactic and extragalactic components,
including those with multiple populations, together with
their property dependencies with respect to the stellar \meta \, and/or \alfaFe \, distributions.

Finally, the next step in building similar future large spectra grids
would be to overcome the assumption adopted when computing
synthetic spectra with extreme \alfaFe \, values not consistent with
the available model atmospheres.
Considering even more complete molecular linelists (including
polyatomic molecules) would also
be of great interest for the coolest spectra of the present grid.
We also point out that
an extension to hotter \T \, with ATLAS12 Kurucz stellar 
atmosphere models is planned for the future, adopting the same linelists (extended towards
higher ionisation stages) and 
consistent solar abundances in order to keep a high consistency between
different spectra grids covering a much larger range of effective temperatures.

\begin{acknowledgements}
The AMBRE project has been supported by the
Centre National d'Etudes Spatiales (CNES), the European Southern
Observatory and the Observatoire de la C\^ote d'Azur (OCA).
The spectrum calculations have been performed with the high-performance computing facility 
SIGAMM, hosted by OCA. We acknowledge the Vienna Atomic Line Database (VALD) for
its compilation of atomic line parameters.  
We sincerely thank B. Edvardsson for his encouraging remarks on this
paper and the stellar atmosphere group
in Uppsala for providing the MARCS model atmospheres to the community.

\end{acknowledgements}

\end{document}